\documentclass[aps,prd,tightenlines,showpacs,amsmath,amssymb]{revtex4}%
\usepackage{graphicx}
\usepackage{epsfig}
\usepackage{graphics,color}
\usepackage{amssymb}
\usepackage{amsmath}
\usepackage{latexsym}
\usepackage{hyperref}
\usepackage{amsbsy}
\usepackage{bm}
\usepackage{url}

\begin{document}
\title{On the symmetry improved CJT formalism in the $O(4)$ linear sigma model}
\author{Hong Mao$^{1,2}$}
\email {mao@hznu.edu.cn}
\address{1.Department of Physics, Hangzhou Normal University, Hangzhou 310036, China \\
2.Theoretical Research Division, Nishina Center, RIKEN, Saitama 351-0198, Japan}


\begin{abstract}
By using the symmetry improved CJT effective formalism developed by Pilaftsis and Teresi, the chiral phase transition is reconsidered in the framework of the $O(4)$ linear sigma model in chiral limit. Our results confirm the restorations of the second-order phase transition and the Goldstone theorem in the Hartree approximation. Finally, we explicitly calculate the effective potentials via the order parameter for various temperatures and address advantages of the present method in comparison with the $O(N)$ model in large-$N$ approximation.
\end{abstract}


\maketitle

\section{Introduction}
It is widely believed that at sufficiently high temperatures and densities there are quantum chromodynamics (QCD) phase transitions between normal hadronic matter and quark-gluon plasma (QGP), where quarks and gluons are no longer confined in hadrons\cite{Rischke:2003mt}\cite{Yagi:2005yb}. Experimentally, the studies of QCD phase transitions are supported by the heavy-ion collisions in laboratories at ultrarelativistic energies, such as the Relativistic Heavy-Ion Collider (RHIC) at Brookhaven National Laboratory, the Large Hadron Collider (LHC) at CERN and future facilities. We believe that these experiments will provide us with a chance to create a hot QCD matter and study its property.

To explore the QCD phase diagram, besides the first principle calculations using lattice techniques (LQCD), there is, however, an alternative approach which respects some salient features of QCD. Following this path, it is necessary to use a variety of effective phenomenological models and QCD-like theories in order to mimic some essential characteristics intimately related to QCD. As a quite popular model, the linear sigma model \cite{GellMann:1960np} for the phenomenology of QCD has been composed to describe the vacuum structure with incorporating chiral symmetry and its spontaneous breaking. The model can be used to describe a restoration of the chiral symmetry at finite temperature in the Hartree-Fock or Hartree approximation\cite{Rischke:2003mt,Roh:1996ek,AmelinoCamelia:1997dd,Petropoulos:1998gt,Lenaghan:1999si,Lenaghan:2000ey,Roder:2003uz,Phat:2003eh,Nemoto:1999qf,Petropoulos:2004bt}
within the Cornwall-Jackiw-Tomboulis (CJT) formalism\cite{Cornwall:1974vz}, also at finite isospin chemical potential\cite{Mao:2006zr,Andersen:2006ys,Shu:2007na,Andersen:2008qk,Phat:2011zz}. However, except the case in the larger-$N$ approximation \cite{AmelinoCamelia:1997dd}\cite{Petropoulos:1998gt}\cite{Lenaghan:1999si}, these kinds of studies suffer from two major drawbacks. One is that, in chiral limit, the CJT effective action violates the Goldstone theorem and gives massive pions in the spontaneous symmetry breaking phase. The other is that the temperature-dependent order parameter predicts a first order phase transition. This prediction, of course, disagrees with the rigorous universality arguments, where the chiral phase transition can be of second order if the $U(1)_A$ symmetry is explicitly broken by instanton for $N_f=2$ flavors of massless quarks \cite{Pisarski:1983ms}. While the former conclusion comes into conflict with recent results reported by Chiku and Hatsuda\cite{Chiku:1997va}\cite{Chiku:1998kd}, where by employing the optimal perturbation theory (OPT) and the real time formalism, they proved that the Goldstone theorem is always satisfied in any given order of the loop expansion in OPT in the $O(4)$ linear sigma model. However, in their works, the application of the OPT nonperturbative method in the model only predicted a first order phase transition in chiral limit, so that the OPT method does not provide a satisfied answer to these two problems yet.

Within the CJT formalism, there are many attempts made to restore the Goldstone theorem in chiral limit, nevertheless, the existing approaches still could not provide a satisfactory solution to this long-standing problem\cite{Nemoto:1999qf, Ivanov:2005yj, Baacke:2002pi,vanHees:2001ik, Marko:2013lxa, Seel:2011ju, Grahl:2011yk}. Very recently, Pilaftsis and Teresi have developed a novel symmetry improved CJT formalism by consistently encoding global symmetries in loop-wise expansions or truncations of the CJT effective action\cite{Pilaftsis:2013xna}. This formalism seems to avoid the pathologies of the existing approaches and satisfy many field-theoretic properties. Taking the $O(2)$ scalar model as simple example, they showed the phase transition is of second order and the Goldstone boson is massless even in the Hartree-Fock approximation. Inspired by their previous work for the $O(2)$ model, in this work, we extend this formalism to more realistic model, the $O(4)$ linear sigma model, to reconsider the chiral phase transition at finite temperature in chiral limit. The advantage of this straightforward extension will become more apparent when we directly apply this discussion to study the formation and evolution of the pion string in early universe and heavy ion collision in thermal background\cite{Zhang:1997is,Mao:2004ym}, where the Goldstone theorem and the features of the second-order phase transition are believed to play essential role\cite{Zurek:1996sj}\cite{Villenkin00}.

The organization of this paper is as follows. In the next section we introduce the model and fix the parameters. In Sect. III, we calculate the effective potential by using the symmetry improved CJT formalism and obtain a set of gap equations. In Sect. IV, the solutions of these gap equations are presented, and the thermal effective masses and potential are calculated. In Sect.V, we briefly review the CJT formulations of the effective potential in an $O(N)$ model in large-$N$ approximation in comparison with the present model. Finally, Section VI summarizes our conclusions and discusses possible future studies.

\section{The Model}
The meson sector Lagrangian of the $SU(2)_{R}\times SU(2)_{L}$ symmetry linear sigma model has the form
\begin{equation}\label{Eq:Lag1}
{\cal L}=\frac{1}{2} \left(\partial _{\mu}\sigma \partial ^{\mu}\sigma +
\partial _{\mu}\vec{\pi} \cdot \partial ^{\mu}\vec{\pi}\right)
-U(\sigma ,\vec{\pi}) ,
\end{equation}
where the potential for the $\sigma$ and $\vec{\pi}$ is parameterized as
\begin{equation}\label{Eq:Potential1}
U(\sigma,
\vec{\pi})=\frac{m^2}{2}(\sigma^2+\vec{\pi}^2)+\frac{\lambda}{24}(\sigma^2+\vec{\pi}^2)^2.
\end{equation}
The $SU(2)_L\times SU(2)_R$ symmetry of the linear sigma model can be
explicitly broken if the potential $U(\sigma, \vec{\pi})$ is made
slightly asymmetric by adding a term $\epsilon \sigma$.
With this addition, the vector isospin $SU(2)$ symmetry remains
exact, but the axial $SU(2)$ transformation is no longer invariant. To the leading order in $\epsilon$, this shifts the minimum of the
potential to $f_{\pi}+\frac{3 \epsilon}{\lambda f_{\pi}^2}$, as a result the pions acquire mass
$m^2_{\pi}=\frac{\epsilon}{f_{\pi}}$.

At tree level and zero temperature the parameters of the Lagrangian
are fixed in a way that these masses agree with the observed value of pion masses and the most commonly accepted value for sigma mass. Unlike the pions, unfortunately, the mass of the sigma meson is still a poorly known number, but the most recent compilation of the Particle Data Group considers that $m_{\sigma}$ can vary from $400$ MeV to $550$ MeV with a full width of $400-700$ MeV\cite{Beringer:1900zz}. So that we take $m_{\sigma}=500$ MeV and $f_{\pi}=93$ MeV as typical values in this work. The coupling constant
$\lambda$ of the model then can be related to zero temperature properties
of the pions and sigma through the expression
\begin{equation}
\lambda =\frac{3(m^2_{\sigma}-m^2_{\pi})}{f^2_{\pi}}.
\end{equation}
The negative mass parameter $m^2$ is introduced in order to ensure
spontaneous breaking of symmetry and its value is chosen to be
\begin{equation}
-m^2=(m^2_{\sigma}-3m^2_{\pi})/2>0~.
\end{equation}

In the chiral limit $\epsilon=0$, the parameters $\lambda$ and $-m^2$ appearing in the Lagrangian can be further simplified as
\begin{equation}\label{lambda}
\lambda =\frac{3 m^2_{\sigma}}{f^2_{\pi}}, \qquad  -2 m^2=m^2_{\sigma}>0~.
\end{equation}
In the following discussion, we will focus on the case of chiral limit and investigate the natural properties of chiral phase transition within the symmetry improved CJT formalism in detail.

\section{The effective potential in the symmetry improved CJT formalism}
In this section we derive the symmetry improved CJT effective
potential for the linear sigma model at finite temperature. The starting point is the Lagrangian given in Eq.~(\ref{Eq:Lag1})
with the choice of the symmetric potential given in
Eq.~(\ref{Eq:Potential1}). By shifting the sigma field as
$\sigma\rightarrow \sigma+\phi$, the classical potential takes the
form
\begin{equation}
U(\phi)=\frac{1}{2}m^2\phi^2+\frac{\lambda}{24}\phi^4.
\end{equation}
The tree-level sigma and pion propagators corresponding to the
above Lagrangian have the form
\begin{subequations}
\begin{eqnarray}
D_{\sigma}^{-1} &=& \omega_{n}^2+\vec{k}^2+m^2+\frac{1}{2}\lambda \phi^2, \\
D_{\pi}^{-1} &=&\omega_{n}^2+\vec{k}^2+m^2+\frac{1}{6}\lambda \phi^2. \label{eq-pro1}
\end{eqnarray}
\end{subequations}
We use the imaginary-time formalism to compute quantities at nonzero temperature, our notation is
\begin{eqnarray}
\int\frac{d^4k}{(2\pi)^4}f(k) \rightarrow \frac{1}{\beta}\sum_n
\int\frac{d^3\vec{k}}{(2\pi)^3}f(i\omega_n,\vec{k}) \nonumber
\equiv \int_{\beta}f(i\omega_n,\vec{k}),
\end{eqnarray}
where $\beta$ is the inverse temperature, $\beta=\frac{1}{k_BT}$,
and as usual Boltzmann's constant is taken as $k_B=1$, and
$\omega_n=2\pi nT$, $n=0, \pm1, \pm2, \pm3, \cdots$. For simplicity we
have introduced a subscript $\beta$ to denote integration and
summation over the Matsubara frequency sums.

Evaluating the effective potential in Hartree approximation means
that one needs to take into account the ``$\infty$" type diagram
only. In the case of the linear sigma model, we can
derive the following expression for the effective potential at
finite temperature\cite{Cornwall:1974vz}
\begin{eqnarray}\label{epotential0}
V(\phi,G_{\sigma}, G_{\pi}) &=& U(\phi)+\frac{1}{2}\int_{\beta}\ln
G^{-1}_{\sigma}(\phi;k)+\frac{3}{2}\int_{\beta}\ln
G^{-1}_{\pi}(\phi;k)
\nonumber\\&&+\frac{1}{2}\int_{\beta}\left[D^{-1}_{\sigma}(\phi;k)G_{\sigma}(\phi;k)-1 \right]
+\frac{3}{2}\int_{\beta}\left[ D^{-1}_{\pi}(\phi;k)G_{\pi}(\phi;k)-1 \right]
\nonumber\\&& +V_2(\phi,
G_{\sigma}, G_{\pi}).
\end{eqnarray}
The first term $U(\phi)$ is the classical
potential and the last term $V_2(\phi, G_{\sigma}, G_{\pi})$
denotes the contribution from the ``$\infty$" diagrams, which is equivalent to the Hartree approximation. Explicitly,
\begin{eqnarray}
V_2(\phi, G_{\sigma}, G_{\pi}) &=& 3\frac{\lambda}{24}
\left[\int_{\beta}G_{\sigma}(\phi;k)
\right]^2+6\frac{\lambda}{24}\left[\int_{\beta}G_{\sigma}(\phi;k)\right]\left[\int_{\beta}G_{\pi}(\phi;k)
\right] \nonumber \\&&
+15\frac{\lambda}{24} \left[\int_{\beta}G_{\pi}(\phi;k)\right]^2.
\end{eqnarray}

Minimizing the effective potential with respect to full
propagators we obtain the following system of
nonlinear gap equations:
\begin{subequations}\label{Eq:Fullpro}
\begin{eqnarray}
G_{\sigma}^{-1} &=&
D_{\sigma}^{-1}+\frac{\lambda}{2}\int_{\beta}G_{\sigma}(\phi;k)
+\frac{\lambda}{2}\int_{\beta}G_{\pi}(\phi;k), \\
G^{-1}_{\pi} &=& D^{-1}_{\pi}+\frac{\lambda}{6}\int_{\beta}G_{\sigma}(\phi;k)
+\frac{5\lambda}{6}\int_{\beta}G_{\pi}(\phi;k).
\end{eqnarray}
\end{subequations}
The bare propagators $D_{\sigma}$ and $D_{\pi}$ are given above. Since the self-energies in Eqs.(\ref{Eq:Fullpro}) are independent of momentum in the Hartree approximation, the full propagators assume the simple form
\begin{eqnarray*}
G_{\sigma}^{-1} &=& \omega_{n}^2+\vec{k}^2+M^2_{\sigma},\\
G_{\pi}^{-1} &=& \omega_{n}^2+\vec{k}^2+M^2_{\pi},
\end{eqnarray*}
where $M_{\sigma}$ and $M_{\pi}$ are the effective masses of
$\sigma$ meson and $\pi$ dressed by interaction
contributions from the ``$\infty$" diagrams. Then the sigma and pion thermal effective masses are given by
\begin{subequations}\label{e-mass}
\begin{eqnarray}
M_{\sigma}^2 &=&
m^2+\frac{\lambda}{2}\phi^2+\frac{\lambda}{2}F(M_{\sigma})
+\frac{\lambda}{2}F(M_{\pi}), \\
M_{\pi}^2 &=&
m^2+\frac{\lambda}{6}\phi^2+\frac{\lambda}{6}F(M_{\sigma})
+\frac{5 \lambda}{6}F(M_{\pi}).
\end{eqnarray}
\end{subequations}
Here we have used a shorthand notation and
introduced the function
\begin{equation}\label{Eq:FM}
F(M)=\int_{\beta}\frac{1}{\omega_{n}^2+\vec{k}^2+M^2}.
\end{equation}
At the level of the Hartree  approximation, the self-energies are momentum independent and thermal effective masses are functions of the order parameter $\phi$ and temperature $T$. The loop integral in equation (\ref{Eq:FM}) requires renormalization \cite{Lenaghan:1999si}. Renormalization in many-body approximation schemes is a nontrivial procedure, and this problem has been extensively discussed in Refs.\cite{vanHees:2001ik,Blaizot:2003an,Berges:2005hc,Fejos:2007ec}. In the following discussions, we follow the ``counter term'' renormalization scheme (CT) adopted in Ref.\cite{Lenaghan:1999si}. As shown in \cite{Marko:2013lxa}\cite{Marko:2012wc}, in the Hartree-Fock approximation, the CT scheme removes the divergences and the continuum limit can be considered without any obstructions in the chiral limit. However, for higher order truncations, a more general renormalization procedure of the CJT functional needs to be employed in order to remove the obstructions in the chiral limit appeared in Ref.\cite{Lenaghan:1999si}.

In general, the effective potential of the CJT formalism can be written down in terms of the solutions of the gap equations (\ref{e-mass}). By minimizing the potential with respect to the expectation value for the sigma field $\phi$, we obtain one more the stationarity equation
\begin{eqnarray}\label{phi}
m^2+\frac{\lambda}{6}\phi^2+\frac{\lambda}{2}F(M_{\sigma})+\frac{\lambda}{2}F(M_{\pi})=0.
\end{eqnarray}
By solving the system of the three equations (\ref{e-mass}) and (\ref{phi}), the effective masses as functions of temperature can be calculated. However, as we know this kind of ``standard" procedure will produce the ``wrong" results, such as the first-order chiral phase transition and the massive Goldstone boson in the model in chiral limit. These difficulties are largely due to the fact that in the loop expansion of the CJT effective action the global symmetries are not exactly maintained at a given loop order of the expansion. In our case, the truncation of the CJT effective action in Hartree approximation violates the Goldstone theorem by higher order terms, which in turn gives rise to massive pions in the spontaneous symmetry breaking phase. In order to resolve these problems, some special efforts should be made in the framework of the usual CJT formalism. For example, in the $O(N)$ model, only that some residual terms are subtracted in the large-$N$ approximation, the model then can predict a second-order phase transition, meanwhile, the Goldstone boson is massless\cite{Petropoulos:1998gt}\cite{Lenaghan:1999si}.

Unlike the CJT effective action, the usual One-particle-irreducible (1PI) action does not suffer from above pathology and respects all global and local symmetries order by order in perturbation theory. Hence, in the 1PI formalism, the Goldstone theorem will retain in any truncation of the 1PI effective action, even such a truncation is loopwise or includes partial resummation of graphs\cite{Chiku:1997va}\cite{Chiku:1998kd}\cite{Duarte:2011ph}. This result is guaranteed by the Ward Identities associated with global symmetries of the theory in the 1PI formalism. In contrast, for the case in the CJT formalism, this could not be applied directly anymore. As we know, the exact all-orders solutions obtained from the gap equations based on the complete CJT effective action could satisfy all Ward Identities of the respective 1PI effective action, since it is believed that the complete CJT effective action is formally equivalent to the complete 1PI effective action \cite{Cornwall:1974vz}. Thus, the Goldstone theorem is fulfilled and the Goldstone bosons are massless in the spontaneous symmetry breaking phase if only the complete CJT effective action is considered. However, as long as the truncations of the CJT effective action are introduced, the approximation solutions satisfy Ward Identity respected to a given truncated CJT effective action would crucially differ from the corresponding Ward Identity derived in the 1PI formalism. Moreover, the former Ward Identity for the truncated CJT effective action does not impose a requirement that the Goldstone bosons should be massless in the Hartree-Fock approximation. As a result, in order to recover the Goldstone theorem in the truncated CJT effective action, in analogy to the standard One-Particle-Irreducible effective action, a properly modified formalism for the CJT effective action has been proposed in Ref.\cite{Pilaftsis:2013xna} through the introduction of a constraint, which has been used to govern the Goldstone theorem in the 1PI effective action. Eventually, in addition to the gap equations (\ref{Eq:Fullpro}), the constraint inserted into the stationarity equations has the form
\begin{eqnarray}\label{phi2}
\phi M_{\pi}^2=0,
\end{eqnarray}
and this constraint can be consistently implemented by redefining the truncated CJT effective action with a Lagrange multiplier field. Once the constrain is included, the stationarity equation (\ref{phi}) derived from the usual CJT formalism certainly needs not to be necessarily obeyed since the CJT effective action or the effective potential has already been modified accordingly as described in Ref.\cite{Pilaftsis:2013xna}.

In the symmetry breaking phase of the model, the constrain (\ref{phi2}) implies $M_{\pi}^2=0$, yielding
\begin{subequations}\label{e-mass2}
\begin{eqnarray}
M_{\sigma}^2 &=&
m^2+\frac{\lambda}{2}\phi^2+\frac{\lambda}{2}F(M_{\sigma})
+\frac{\lambda}{2}F(0), \\
0 &=&
m^2+\frac{\lambda}{6}\phi^2+\frac{\lambda}{6}F(M_{\sigma})
+\frac{5 \lambda}{6}F(0).
\end{eqnarray}
\end{subequations}
Now the massless pions are naturally realized whenever the order parameter is not vanished. On the contrary, in the symmetric phase, we have $\phi=0$ and the constraint is automatically satisfied. As a consequence, the two mass-gap equations become degenerate, the particles have the same mass with $M_{\sigma}=M_{\pi}\equiv M$, and
\begin{equation}\label{Eq:Gap}
M^2=m^2+\lambda F(M).
\end{equation}
It is worth to point out that the constraint proposed in Eq.(\ref{phi2}) can be obtained without any assumption in the large-$N$ approximation in the framework of the usual CJT formalism, therefore, the large-$N$ approximation in the $O(N)$ model implies that the pions should be massless in the symmetry broken phase also. In other words, once we take the constraint(\ref{phi2}), the well-known results in the large-$N$ approximation could be copied.

Using the definition in the framework of the symmetry improved CJT formalism, we can rewrite the finite temperature effective potential $V_{eff}$ as a function of $\varphi$ in the thermal Hartree approximation by extending the mass-gap equations (\ref{e-mass}) from $\phi \mapsto \varphi$ while
\begin{equation}\label{potential}
M_{\pi}^2 (\varphi)=\frac{1}{\varphi} \frac{d V_{eff}(\varphi)}{d \varphi}.
\end{equation}
Now the Goldstone boson masslessness condition can be naturally implemented in the symmetry breaking phase, and the integration of the effective potential $V_{eff}$ with respect to the field $\varphi$ really create a typical potential possessing the second order phase transition. As mentioned in above, an explicit symmetry breaking term $\epsilon \sigma$ can be included into the original Lagrangian to generate the pion observed masses. Since the $\epsilon $ only appears in the classical potential, rather than in the gap equations, therefore the introduction of this term in our discussions would be rather trivial by adding additional term $\epsilon \varphi$ in the effective potential $V_{eff}$ in Eq.(\ref{potential}). Accordingly, the constraint is rewritten as
\begin{eqnarray}\label{phi3}
\varphi M_{\pi}^2=\epsilon.
\end{eqnarray}
Then at low temperature the pions appear with the observed mass since the global symmetry has already been explicitly broken.

\section{Numerical results}

In this section, we discuss the numerical results at non-zero temperature in chiral limit. We solve the system of mass-gap Eqs.(\ref{e-mass}) with the constraint (\ref{phi2})
using a numerical method based on the Newton-Raphson method of solving nonlinear equations. In this way, we are able to determine the effective masses $M_{\sigma}$ and $M_{\pi}$ and the order parameter $\phi$ as functions of temperature $T$. For comparison with the previous studies, we also solve the gap equations with the stationarity equation (\ref{phi}) according to the usual CJT scenario.

As mentioned in previous section, the loop integral in equation (\ref{Eq:FM}) is divergent and requires renormalization. In the CT scheme, the UV divergences in $F(M)$ can be removed by the counter terms as illustrated in Ref.\cite{Lenaghan:1999si}. Similarly to \cite{Lenaghan:1999si}, it is an easy exercise to write down the renormalized gap equations by replacing $F(M)$ with
\begin{eqnarray}
F(M) =F_T(M) +F_{\mu}(M),
\end{eqnarray}
where
\begin{eqnarray}
F_T(M)=\int\frac{d^3\vec{k}}{(2\pi)^3}\frac{1}{\omega_{\vec{k}}}\frac{1}{e^{\beta\omega_{\vec{k}}}-1}
\end{eqnarray}
with $\omega_{\vec{k}}\equiv\sqrt{\vec{k}^2+M^2}$, and
\begin{eqnarray}
F_{\mu}(M)\equiv \frac{1}{(4 \pi)^2}\left[ M^2 \ln \frac{M^2}{\mu^2}-M^2+\mu^2 \right].
\end{eqnarray}
The renormalization scale $\mu$ is chosen to give the correct values for sigma and pion masses at vacuum.  In Hartree approximation, the gap equations (\ref{e-mass}) and (\ref{e-mass2})in chiral limit at vacuum give
\begin{eqnarray}
m_{\sigma}^2=\frac{\lambda f_{\pi}^2}{3}+\frac{\lambda}{3}\frac{m_{\sigma}^2}{16 \pi^2} \ln \left[ \frac{m_{\sigma}^2}{\mu^2 e} \right],
\end{eqnarray}
where we take the pion to be a massless meson and $\phi=f_{\pi}$ at $T=0$ MeV. In order to be consistent with the equation (\ref{lambda}), there is only a single choice for the renormalization scale, $\mu^2\equiv m_{\sigma}^2/e$. Then the negative mass parameter $m^2$ is given by
\begin{eqnarray}
m^2=-\frac{m_{\sigma}^2}{2}-\lambda \frac{\mu^2}{16 \pi^2}.
\end{eqnarray}

\begin{figure}[thbp]
\epsfxsize=9.0 cm \epsfysize=6.5cm
\epsfbox{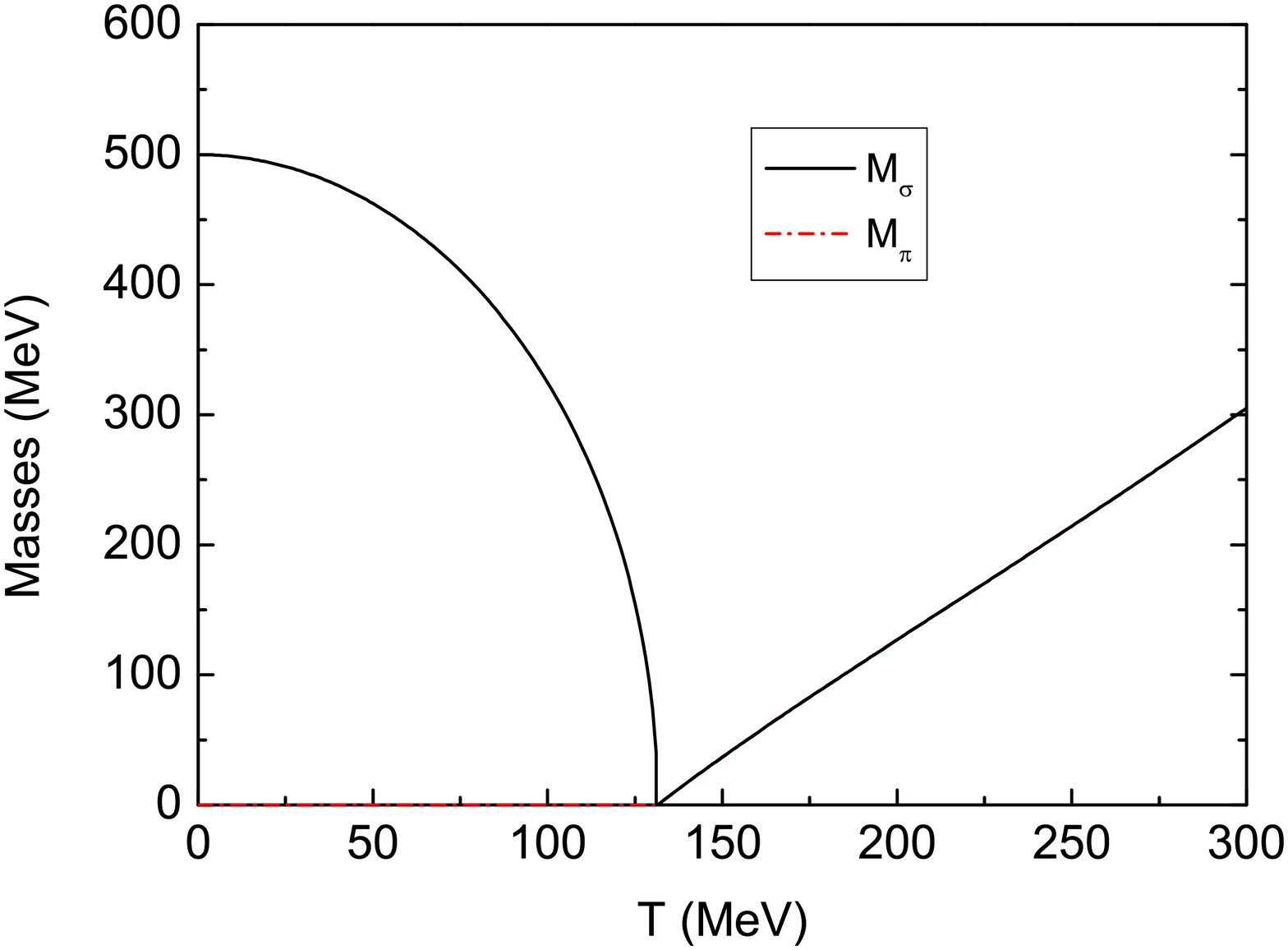}\hspace*{0.1cm} \epsfxsize=9.0 cm
\epsfysize=6.5cm \epsfbox{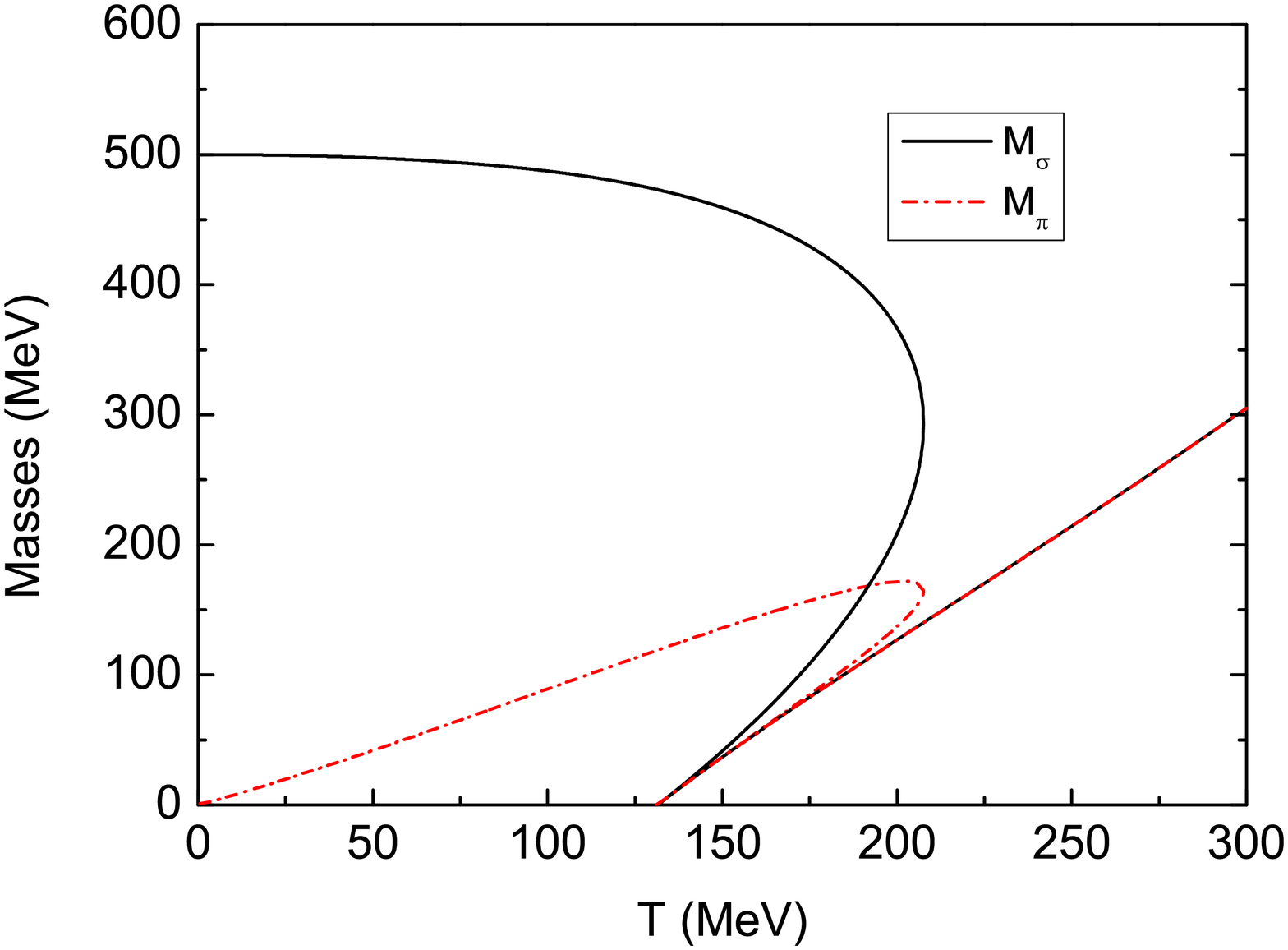}
\vskip -0.05cm \hskip 0.15 cm \textbf{( a ) } \hskip 6.5 cm \textbf{( b )} \\
 \caption{Solution of the system of gap equations in
the case of chiral limit. (a) The thermal effective masses of the sigma and pions are given as functions of temperature within the symmetry improved CJT formalism. (b)The thermal effective masses of the sigma and pions are given as functions of temperature within the traditional CJT formalism. }
\label{Fig01-02}
\end{figure}

With the above fixed renormalization scale, the sigma and pion masses for various temperature $T$ are shown in Fig.\ref{Fig01-02} within the symmetry improved CJT formalism and the usual CJT formalism. In both formalisms the sigma mass first smoothly decreases and then rebounds and grows again at high $T$. At large $T$ all mesons masses grow linearly with increasing $T$. For the pion masses, at low temperature, in the case of the symmetry improved CJT formalism, they appear as massless Goldstone bosons until the temperature $T$ approaches some critical temperature $T_c$ around $131.5$ MeV. After that, the thermal contribution to the effective masses make them degenerate with the sigma signaling a restoration of chiral symmetry. From the right panel in Fig.\ref{Fig01-02}, the thermal effective masses of sigma and of pions just show the typical behaviors of the first-order phase transition within the usual CJT formalism.

The critical temperature $T_c$ can be calculated in high temperature limit where we can set $\phi=0$ and $M_{\pi}=M_{\sigma}$, then we only need to solve just one equation (\ref{Eq:Gap}). If we further decrease the temperature but keep the $\phi=0$, we can reach some point where the mass of the particles vanishes, then this equation reduces to a well known result
\begin{equation}
0=-\frac{m_{\sigma}^2}{2}+\lambda \frac{T^{2}}{2 \pi^{2}}\frac{\pi^2}{6}.
\end{equation}
This relationship exactly defines the transition temperature $T_{c}=\sqrt{2} f_{\pi}\simeq 131.5$ MeV in the model within both formalisms, in which all particle masses become zero. Can this point be also realized if we go from the low temperature to high temperature too? For the case of the usual CJT formalism, by combining the gap equations (\ref{e-mass}) with the stationarity equation (\ref{phi}), we arrive at
\begin{equation}
M_{\sigma}^2=\frac{1}{3}\lambda \phi^2.
\end{equation}
This equation guarantees the sigma mass varies proportionally to the order parameter, and for $\phi=0$, we always observe $M_{\sigma}=0$. Similarly, for the case of the symmetry improved CJT formalism, we can eliminate $\phi$ in the equations (\ref{e-mass2}) to end up with the following equation
\begin{equation}\label{sigmamass}
M_{\sigma}^2=-2 \left( m^2+\lambda F(0) \right)=\frac{\lambda}{6}\left( T_c^2-T^2 \right).
\end{equation}
The last equation leads to an obvious output $M_{\sigma}=M_{\pi}=0$ for $T=T_c$, and the masses of sigma meson in the symmetry breaking phase is independent of the renormalization scale.

\begin{figure}
\includegraphics[scale=0.36]{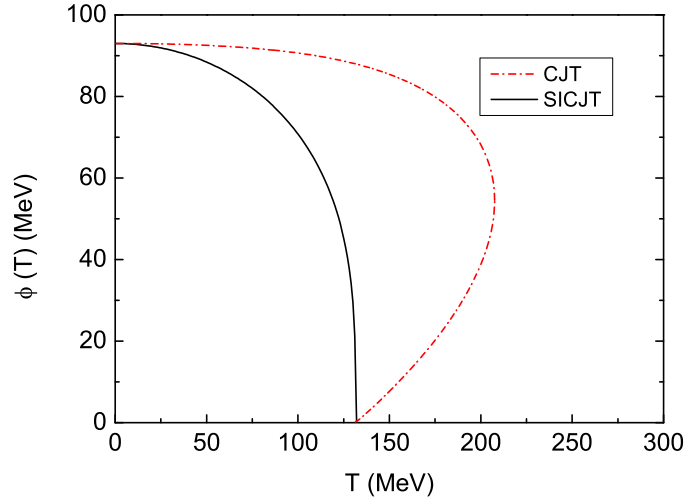}
\caption{\label{Fig03} Evolution of the order parameter $\phi$ as a function of $T$ in the Hartree approximation with the symmetry improved CJT formalism (SICJT) and the traditional CJT formalism (CJT) in chiral limit.}
\end{figure}

\begin{figure}
\includegraphics[scale=0.36]{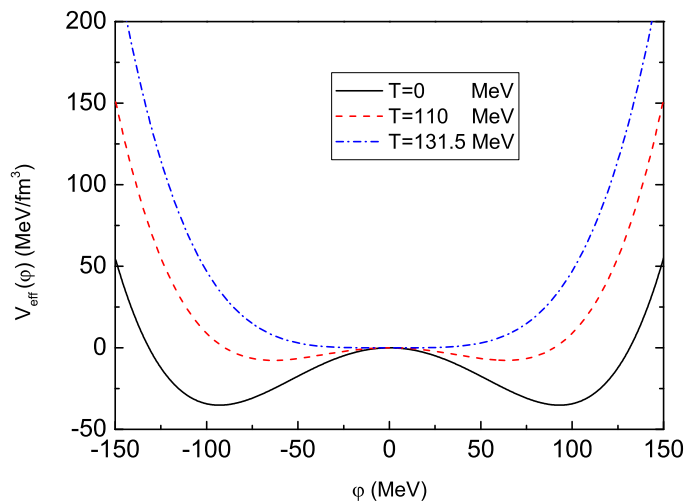}
\caption{\label{Fig04} Symmetry improved CJT effective potential $V_{eff}$ in Hartree approximation as a function of the order parameter $\varphi$ for several temperatures: $T=0$ MeV, $T=110$ MeV and $T=131.5$ MeV.}
\end{figure}

The order parameter $\phi$ vanishes continuously in the symmetry improved CJT formalism case is shown in Fig.\ref{Fig03}, this could be taken as a corroborator to show that there is a second-order phase transition in the Hartree approximation within the symmetry improved CJT formalism. To display this feature more explicitly, we can calculate the effective potential $V_{eff}$ as a function of the temperature and the order parameter by integrating the effective potential $V_{eff}$ with respect to the field $\varphi$. Fig.\ref{Fig04} presents the numerical solution of the equation (\ref{potential}) at three different temperatures, $T=0$ MeV, $T=110$ MeV and $T=131.5$ MeV. The shape of the potential tells that a second-order phase transition takes place in the model in analogy to the standard 1PI effective action.

\section{Comparison with large-$N$ approximation}

From the above discussions, only the large-$N$ limit in the $O(N)$ model, a consistent prediction in the usual CJT formalism is obtained, where the chiral phase transition is of second order and the Goldstone boson is massless. In this section, we first briefly review the basic structure of the $O(N)$ linear sigma model\cite{Petropoulos:1998gt}\cite{Lenaghan:1999si} so that we can discuss and reveal the possible differences and advantages of the symmetry improved CJT formalism presented in the previous section.

The $O(4)$ linear sigma model can be promoted to a more general version, the $O(N)$ model or vector model, based on a set of $N$ real scalar fields. Its Lagrangian is given by\cite{Petropoulos:1998gt}\cite{Lenaghan:1999si}
\begin{equation}\label{Eq:Lag2}
{\cal L}_{N}=\frac{1}{2} \left(\partial _{\mu}\mathbf{\Phi} \right)^2-\frac{1}{2}m^2 \mathbf{\Phi}^2-\frac{1}{6N}\lambda \mathbf{\Phi}^4+ \epsilon \sigma,
\end{equation}
where $\mathbf{\Phi}=(\sigma,\pi_1,\cdots,\pi_{N-1})$ is an $N$-component scalar field. For $\epsilon=0$ and $m^2>0$, the Lagrangian is invariant under $O(N)$ rotations of the fields, while, for $\epsilon=0$ and $m^2<0$, this symmetry is spontaneously broken down to $O(N-1)$, with $N-1$ Goldstone bosons (the pions).The phenomenological explicit symmetry breaking term, $\epsilon$, is introduced to generate the observed nonzero pion masses. Spontaneous symmetry breaking leads to a non-vanishing vacuum expectation value for sigma field, $<\sigma>=\phi$. In analogous steps in the Hartree case in the last section, by shifting the sigma field as $\sigma\rightarrow \sigma+\phi$, the inverse tree-level sigma and pion propagators are given by
\begin{subequations}
\begin{eqnarray}
D_{\sigma}^{-1} &=& \omega_{n}^2+\vec{k}^2+m^2+\frac{2}{N}\lambda \phi^2, \\
D_{\pi}^{-1} &=&\omega_{n}^2+\vec{k}^2+m^2+\frac{2}{3N}\lambda \phi^2. \label{eq-pro2}
\end{eqnarray}
\end{subequations}
Then the CJT effective potential for the $O(N)$ model at finite temperature is obtained as\cite{Petropoulos:1998gt}\cite{Lenaghan:1999si}
\begin{eqnarray}\label{potential2}
V(\phi,G_{\sigma}, G_{\pi}) &=& \frac{1}{2}m^2 \phi^2+\frac{1}{6N}\lambda \phi^4-\epsilon \sigma+\frac{1}{2}\int_{\beta}\ln
G^{-1}_{\sigma}(\phi;k)+\frac{N-1}{2}\int_{\beta}\ln
G^{-1}_{\pi}(\phi;k)
\nonumber\\&&+\frac{1}{2}\int_{\beta}\left[D^{-1}_{\sigma}(\phi;k)G_{\sigma}(\phi;k)-1 \right]
+\frac{N-1}{2}\int_{\beta}\left[ D^{-1}_{\pi}(\phi;k)G_{\pi}(\phi;k)-1 \right]
\nonumber\\&& +V_2(\phi, G_{\sigma}, G_{\pi}),
\end{eqnarray}
where $V_2(\phi, G_{\sigma}, G_{\pi})$ denotes the contribution from the double-bubble diagrams. In Hartree approximation, its contribution is
\begin{eqnarray}
V_2(\phi, G_{\sigma}, G_{\pi}) &=& 3\frac{\lambda}{6N}
\left[\int_{\beta}G_{\sigma}(\phi;k)
\right]^2+2(N-1)\frac{\lambda}{6N}\left[\int_{\beta}G_{\sigma}(\phi;k)\right]\left[\int_{\beta}G_{\pi}(\phi;k)
\right] \nonumber \\&&
+(N^2-1)\frac{\lambda}{6N} \left[\int_{\beta}G_{\pi}(\phi;k)\right]^2.
\end{eqnarray}
It is easy to check that for $N=4$, we can naturally recover the previous results in the $O(4)$ linear sigma model.

Following the standard procedure\cite{Petropoulos:2004bt}, the stationarity conditions are decided by minimizing the effective potential $V(\phi,G_{\sigma}, G_{\pi})$ in Eq.(\ref{potential2}) with respect to the full propagators $G_{\sigma}$, $G_{\pi}$ and the chiral condensate $\phi$. Then, the dressed sigma and pion masses are determined by the following gap equations:
\begin{subequations}\label{e-mass3}
\begin{eqnarray}
M_{\sigma}^2 &=&
m^2+\frac{2\lambda}{N}\phi^2+\frac{2\lambda}{N}F(M_{\sigma})
+\frac{2(N-1)\lambda}{3N}F(M_{\pi}), \\
M_{\pi}^2 &=&
m^2+\frac{2\lambda}{3N}\phi^2+\frac{2\lambda}{3N}F(M_{\sigma})
+\frac{2(N+1) \lambda}{3N}F(M_{\pi}),
\end{eqnarray}
\end{subequations}
and $\phi$ is determined by
\begin{eqnarray}\label{phi3}
\epsilon = \left(m^2+\frac{2\lambda}{3N}\phi^2+\frac{2\lambda}{N}F(M_{\sigma})+\frac{2(N-1)\lambda}{3N}F(M_{\pi}) \right) \phi.
\end{eqnarray}

In the large-$N$ approximation, one simply neglects all contributions of order $1/N$, therefore, the system of the last three equations reduces to
\begin{subequations}\label{e-mass4}
\begin{eqnarray}
M_{\sigma}^2 &=&
m^2+\frac{2\lambda}{N}\phi^2+\frac{2\lambda}{3}F(M_{\pi}), \\
M_{\pi}^2 &=&
m^2+\frac{2\lambda}{3N}\phi^2+\frac{2\lambda}{3}F(M_{\pi}), \\
\epsilon &=& \left(m^2+\frac{2\lambda}{3N}\phi^2+\frac{2\lambda}{3}F(M_{\pi}) \right) \phi .
\end{eqnarray}
\end{subequations}
This leads to
\begin{subequations}\label{e-mass5}
\begin{eqnarray}
M_{\sigma}^2 &=&
M^2_{\pi}+\frac{4\lambda}{3N}\phi^2, \\
\epsilon  &=& M_{\pi}^2 \phi.
\end{eqnarray}
\end{subequations}
The latter equation implies that in chiral limit, $\epsilon=0$, the large-$N$ approximation gives that the pions should be massless in the low-temperature phase in accordance with the Goldstone theorem. In contrast, at very high temperatures, the potential has only one minimum, $\phi=0$, we again have $M_{\sigma}=M_{\pi}$. In this case,
\begin{eqnarray}
M^2_{\sigma}=M^2_{\pi}=m^2+\frac{2\lambda}{3}F(M_{\pi}),
\end{eqnarray}
the last equation actually defines the critical temperature for the chiral phase transition in the large-$N$ approximation. To evaluate the critical temperature $T_c$ of the chiral phase transition, we set $M_{\sigma}=M_{\pi}=0$ in the last equation and use the well-known results $F_T(0)=T^2/12$ and $m^2=-{m_{\sigma}^2}/{2}-{ \lambda}{\mu^2}/{ (24 \pi^2)}$, after that we find that the critical temperature is at
\begin{eqnarray}\label{critemp}
T'_c=\sqrt{3}\left(\frac{3 m_{\sigma}^2}{\lambda}\right)=\sqrt{3} f_{\pi}\approx 161 \mathrm{MeV}.
\end{eqnarray}
This is different from the case in the symmetry improved CJT formalism, where the critical temperature is set by the equation (\ref{Eq:Gap}) with $T_{c}=\sqrt{2} f_{\pi}\simeq 131.5$ MeV. In order to analyze the order of phase transition, in the spontaneous symmetry breaking phase, that is $\phi\neq 0$, from the gap equations (\ref{e-mass4}) and the relationship in Eqs.(\ref{e-mass5}), the chiral condensate $\phi$ can be described as
\begin{equation}\label{critemp2}
M_{\sigma}^2=\frac{4 \lambda}{3N}\phi^2=-2 \left( m^2+\frac{2\lambda}{3} F(0) \right)=\frac{\lambda}{9}\left( {T'}_c^{2}-T^2 \right).
\end{equation}
Thus, we can conclude that the order parameter $\phi$ vanishes continuously and the chiral phase transition in the large-$N$ approximation corresponds to a second-order phase transition. It is worthy to point out that the results presented in Eqs.(\ref{critemp}) and (\ref{critemp2}) are independent of the renormalization scale $\mu$ choosing, although, now $\mu$ is a free parameter. This also suggests that there is no difference between the unrenormalized and renormalized cases in the phase of spontaneously broken symmetry in the large-$N$ approximation. In comparison with the results demonstrated in the last section, both methods display a qualitative behaviors at finite temperature, except for the critical temperature $T_c$ has been moved towards to a high one, $T'_c$. The reason for such a change is largely due to the fact that the sigma contribution is ignored in the large-$N$ approximation, but reserving in the symmetry improved CJT formalism.

Now, we can find that both methods predict a correct second-order phase transition and reserve the Goldstone theorem as well even in the level of the Hartree approximation. However, in comparison, we can find that the constraint proposed in Eq.(\ref{phi2}) can be automatically realized in the case of large-$N$ approximation in the $O(N)$ linear sigma model\cite{Petropoulos:1998gt}\cite{Lenaghan:1999si}, whereas, it should be artificially introduced in the framework of the symmetry improved CJT formalism\cite{Pilaftsis:2013xna}. Moreover, the stationarity equation in Eq.(\ref{phi}) and the constraint (\ref{phi2}) do not satisfy simultaneously in the case of the symmetry improved CJT formalism in the $O(4)$ linear sigma model, this implies that the effective potential in Eq.(\ref{epotential0}) is inevitably needed to be reconstructed according to the constraint (\ref{phi2}) when leaving the gap equations (\ref{Eq:Fullpro}) untouched. This is the underlying reason why we have to redefine the effective potential $V_{eff}$ as in the equation (\ref{potential}). In this sense, the large-$N$ approximation in the $O(N)$ model seems more self-consistent and advantageous than that of the symmetry improved CJT formalism. Unfortunately, it is not so obvious as it appears if we check the gap equations in both cases. From Eqs.(\ref{e-mass2}) and (\ref{e-mass4}), the sigma contribution is reserved in the case of the symmetry improved CJT formalism, however, it is ignored in the case of large-$N$ approximation. More specifically, in the large-$N$ approximation, only the pionic double-bubble diagrams are actually summed in the effective potential (\ref{potential2}), such a consideration is of course incomplete and lack of self-consistent, and this will bring on the incomplete gap equations, which in turn induce errors when we calculate the critical temperature, especially when we take $N=4$ as a special phenomenological example. Moreover, in the large-$N$ expansion, the Goldstone theorem is satisfied only at the leading order, unless the external propagator is used, the Goldstone bosons become massive within quantum loops, while the Goldstone theorem holds to an arbitrary level of truncation in the symmetry improved formalism. So that the method based on the symmetry improved CJT formalism has partially fixed these crucial flaws in the large-$N$ approximation of the $O(N)$ model and contains more of the QCD phenomenology. These can be treated as apparent advantages of this new CJT scheme.

In the end of this section, it should be noted that from the equations (\ref{sigmamass}) and (\ref{critemp2}), the sigma mass is only dependent on the critical temperature or the sigma mass at vacuum, when submitting these equations into their gap equations, e.g. Eqs.(\ref{e-mass2}) for the symmetry improved CJT formalism and Eqs.(\ref{e-mass4}) for the large-$N$ approximation, $\phi$ can be explicitly expressed as the function of the temperature $T$, $m$ and $M_{\sigma}$. This means that for both methods there is no self-consistent equations to be determined in the symmetry breaking phase, or the mass-gap equations could be solved analytically\cite{Pilaftsis:2013xna}. The inconsistency cannot be avoided once the pion is set as a massless Goldstone boson in the symmetry breaking phase.

\section{Summary and discussion}
We have discussed the effective masses of the mesons and the effective potential at finite temperature in the framework of the $O(4)$ linear sigma model in chiral limit by adopting the symmetry improved CJT formalism. Unlike the traditional CJT formalism, in the Hartree approximation, we find a second-order phase transition, this observation seems to be in agreement with the renormalization group argument: the chiral phase transition in $N_f=2$ QCD is likely to be of second order at $\mu_B=0$\cite{Pisarski:1983ms}. Moreover, we have shown that the Goldstone bosons resulting from spontaneous symmetry breaking of the $O(4)$ symmetry are massless, and a naive truncation of the symmetry improved CJT effective action does not violate the Goldstone theorem in the Hartree approximation. To illustrate the more apparent nature of the second-order phase transition, we also have constructed the thermal effective potential for several temperatures. From the comparison with the results in the usual CJT formalism, we conclude that the symmetry improved CJT formalism really quite improves the original one in the $O(4)$ linear sigma model.

Since the symmetry improved CJT formalism naturally cures the two major problems of the naive CJT method, namely, the breakdown of the Goldstone theorem in symmetry-broken phase and the existence
of first-order phase transition, so that this method is superior than the OPT method\cite{Chiku:1997va}\cite{Chiku:1998kd}, which merely resolve the first problem. Moreover, unlike the $O(N)$ model in large-$N$ approximation\cite{Petropoulos:1998gt}\cite{Lenaghan:1999si}, the sigma contribution in the resummation is kept and included in calculation of the gap equations for the effective masses, hence results evaluated in the present method are believed to be closer to phenomenology related to the real problem of QCD. All these can be taken as apparent advantages over other resummation methods proposed so far.

Of course, in the present study, we have only simply considered the Hartree approximation. Any attempts to go beyond the usual Hartree approximation are worthy to do in order to identify their effects in this novel CJT formalism, e.g. the sunset-type diagrams\cite{Marko:2013lxa}\cite{Roder:2005vt} or more higher loops. Such an extension is straightforward but technically complicated. Finally, the $O(4)$ linear sigma model with two quarks could be combined with the Polyakov loop which allows to investigate both the chiral and the deconfinement phase transition\cite{Schaefer:2007pw}\cite{Mao:2009aq}\cite{Schaefer:2009ui}, and recently in the two-particle irreducible (2PI) approximation, the QCD phase diagram has been already investigated in Ref\cite{Marko:2010cd}. Since here we have the gauge fields and the spontaneous breaking via the quark condensate, it would be more interesting to explore how the symmetry improved CJT formalism works in the Polykov-loop extended quark meson model with two or three flavors by directly comparing with the Lattice data or the other effective models. All these studies will make us getting closer to the real QCD world.

\begin{acknowledgments}
We thank Jinshuang Jin and Nicholas Petropoulos for reading the original manuscript, and Gergely Fejos and Song Shu for valuable comments and discussions. The project is Supported by the Program for Excellent Young Teachers in Hangzhou Normal University, NSFC under Nos.11274085 and 11275002.
\end{acknowledgments}


\begin{thebibliography}{199}

\bibitem{Rischke:2003mt}
  D.~H.~Rischke,
  Prog.\ Part.\ Nucl.\ Phys.\  {\bf 52}, 197 (2004)
  [arXiv:nucl-th/0305030].

\bibitem{Yagi:2005yb}
  K.~Yagi, T.~Hatsuda and Y.~Miake,
  ``\textit{Quark-gluon plasma: From big bang to little bang},''
  Camb.\ Monogr.\ Part.\ Phys.\ Nucl.\ Phys.\ Cosmol.\  {\bf 23}, 1 (2005).  

\bibitem{GellMann:1960np}
  M.~Gell-Mann and MLevy,
  Nuovo Cim.\  {\bf 16}, 705 (1960).  

\bibitem{Roh:1996ek}
  H.~-S.~Roh and T.~Matsui,
  Eur.\ Phys.\ J.\ A {\bf 1}, 205 (1998)
  [nucl-th/9611050].

\bibitem{AmelinoCamelia:1997dd}
  G.~Amelino-Camelia,
  Phys.\ Lett.\ B {\bf 407}, 268 (1997)
  [hep-ph/9702403].

\bibitem{Petropoulos:1998gt}
  N.~Petropoulos,
  J.\ Phys.\ G {\bf 25}, 2225 (1999).

\bibitem{Lenaghan:1999si}
  J.~T.~Lenaghan and D.~H.~Rischke,
  J.\ Phys.\ G {\bf 26}, 431 (2000).

\bibitem{Lenaghan:2000ey}
  J.~T.~Lenaghan, D.~H.~Rischke and J.~Schaffner-Bielich,
  Phys.\ Rev.\ D {\bf 62}, 085008 (2000).

\bibitem{Roder:2003uz}
  D.~Roder, J.~Ruppert and D.~H.~Rischke,
  Phys.\ Rev.\ D {\bf 68}, 016003 (2003).

\bibitem{Phat:2003eh}
  T.~H.~Phat, N.~T.~Anh and L.~V.~Hoa,
  Eur.\ Phys.\ J.\ A {\bf 19}, 359 (2004)
  [hep-ph/0309055].

\bibitem{Nemoto:1999qf}
  Y.~Nemoto, K.~Naito and M.~Oka,
  Eur.\ Phys.\ J.\ A {\bf 9}, 245 (2000)
  [hep-ph/9911431].

\bibitem{Petropoulos:2004bt}
  N.~Petropoulos,
  arXiv:hep-ph/0402136 and references therein.

\bibitem{Cornwall:1974vz}
  J.~M.~Cornwall, R.~Jackiw and E.~Tomboulis,
  Phys.\ Rev.\ D {\bf 10}, 2428 (1974).

\bibitem{Mao:2006zr}
  H.~Mao, N.~Petropoulos and W.~-Q.~Zhao,
  J.\ Phys.\ G {\bf 32}, 2187 (2006)
  [hep-ph/0606241].

\bibitem{Andersen:2006ys}
  J.~O.~Andersen,
  Phys.\ Rev.\ D {\bf 75}, 065011 (2007)
  [hep-ph/0609020].

\bibitem{Shu:2007na}
  S.~Shu and J.~-R.~Li,
  J.\ Phys.\ G {\bf 34}, 2727 (2007)
  [hep-ph/0702230].

\bibitem{Andersen:2008qk}
  J.~O.~Andersen and T.~Brauner,
  Phys.\ Rev.\ D {\bf 78}, 014030 (2008)
  [arXiv:0804.4604 [hep-ph]].

\bibitem{Phat:2011zz}
  T.~H.~Phat and N.~V.~Thu,
  Eur.\ Phys.\ J.\ C {\bf 71}, 1810 (2011).

\bibitem{Pisarski:1983ms}
  R.~D.~Pisarski and F.~Wilczek,
  Phys.\ Rev.\ D {\bf 29}, 338 (1984).

\bibitem{Chiku:1997va}
  S.~Chiku and T.~Hatsuda,
  Phys.\ Rev.\ D {\bf 57}, 6 (1998)
  [hep-ph/9706453].

\bibitem{Chiku:1998kd}
  S.~Chiku and T.~Hatsuda,
  Phys.\ Rev.\ D {\bf 58}, 076001 (1998)
  [hep-ph/9803226].

\bibitem{Ivanov:2005yj}
  Y.~.B.~Ivanov, F.~Riek and J.~Knoll,
  Phys.\ Rev.\ D {\bf 71}, 105016 (2005)
  [hep-ph/0502146];  Y.~.B.~Ivanov, F.~Riek, H.~van Hees and J.~Knoll,
  Phys.\ Rev.\ D {\bf 72}, 036008 (2005)
  [hep-ph/0506157].

\bibitem{Baacke:2002pi}
  J.~Baacke and S.~Michalski,
  Phys.\ Rev.\ D {\bf 67}, 085006 (2003)
  [hep-ph/0210060].

\bibitem{vanHees:2001ik}
  H.~van Hees and J.~Knoll,
  Phys.\ Rev.\ D {\bf 65}, 025010 (2002); {\bf 65}, 105005 (2002); {\bf 66}, 025028 (2002).

\bibitem{Marko:2013lxa}
  G.~Markó, U.~Reinosa and Z.~Szép,
  Phys.\ Rev.\ D {\bf 87}, 105001 (2013)
  [arXiv:1303.0230 [hep-ph]].

\bibitem{Seel:2011ju}
  E.~Seel, S.~Struber, F.~Giacosa and D.~H.~Rischke,
  Phys.\ Rev.\ D {\bf 86}, 125010 (2012)
  [arXiv:1108.1918 [hep-ph]].

\bibitem{Grahl:2011yk}
  M.~Grahl, E.~Seel, F.~Giacosa and D.~H.~Rischke,
  Phys.\ Rev.\ D {\bf 87}, 096014 (2013)
  [arXiv:1110.2698 [nucl-th]].

\bibitem{Pilaftsis:2013xna}
  A.~Pilaftsis and D.~Teresi,
  Nucl.\  Phys.\ B {\bf 874}, 594 (2013)
  [arXiv:1305.3221 [hep-ph]].

\bibitem{Zhang:1997is}
  X.~Zhang, T.~Huang and R.~H.~Brandenberger,
  Phys.\ Rev.\ D {\bf 58}, 027702 (1998)
  [hep-ph/9711452].

\bibitem{Mao:2004ym}
  H.~Mao, Y.~Li, M.~Nagasawa, X.~-m.~Zhang and T.~Huang,
  Phys.\ Rev.\ C {\bf 71}, 014902 (2005)
  [hep-ph/0404132].

\bibitem{Zurek:1996sj}
  W.~H.~Zurek,
  Phys.\ Rept.\  {\bf 276}, 177 (1996)
  [cond-mat/9607135].

\bibitem{Villenkin00} A. Villenkin, E.P.S. Shellard, {\em Cosmic strings
and Other topological defects } (Cambridge University Press,
Cambridge, 2000).

\bibitem{Beringer:1900zz}
  J.~Beringer {\it et al.}  [Particle Data Group Collaboration],
  Phys.\ Rev.\ D {\bf 86}, 010001 (2012).  

\bibitem{Blaizot:2003an}
  J.~-P.~Blaizot, E.~Iancu and U.~Reinosa,
  Nucl.\ Phys.\ A {\bf 736}, 149 (2004)
  [hep-ph/0312085].

\bibitem{Berges:2005hc}
  J.~Berges, S.~Borsanyi, U.~Reinosa and J.~Serreau,
  Annals Phys.\  {\bf 320}, 344 (2005)
  [hep-ph/0503240].

\bibitem{Fejos:2007ec}
  G.~Fejos, A.~Patkos and Z.~.Szep,
  Nucl.\ Phys.\ A {\bf 803}, 115 (2008)
  [arXiv:0711.2933 [hep-ph]].

\bibitem{Marko:2012wc}
  G.~Marko, U.~Reinosa and Z.~Szep,
  Phys.\ Rev.\ D {\bf 86}, 085031 (2012)
  [arXiv:1205.5356 [hep-ph]].

\bibitem{Duarte:2011ph}
  D.~C.~Duarte, R.~L.~S.~Farias and R.~O.~Ramos,
  Phys.\ Rev.\ D {\bf 84}, 083525 (2011)
  [arXiv:1108.4428 [hep-ph]].

\bibitem{Roder:2005vt}
  D.~Roder, J.~Ruppert and D.~H.~Rischke,
  Nucl.\ Phys.\ A {\bf 775}, 127 (2006)
  [hep-ph/0503042].

\bibitem{Schaefer:2007pw}
  B.~J.~Schaefer, J.~M.~Pawlowski and J.~Wambach,
  Phys.\ Rev.\  D {\bf 76}, 074023 (2007).

\bibitem{Mao:2009aq}
  H.~Mao, J.~Jin and M.~Huang,
  J.\ Phys.\ G {\bf 37}, 035001 (2010).  

\bibitem{Schaefer:2009ui}
  B.~-J.~Schaefer, M.~Wagner and J.~Wambach,
  Phys.\ Rev.\ D {\bf 81}, 074013 (2010).  


\bibitem{Marko:2010cd}
  G.~Marko and Z.~.Szep,
  Phys.\ Rev.\ D {\bf 82}, 065021 (2010)
  [arXiv:1006.0212 [hep-ph]].

\end{thebibliography}
\end{document}